\documentclass[fleqn,twoside]{article}
\usepackage{espcrc2}
\usepackage{graphicx}

\title{Measurements of Cherenkov Photons with Silicon Photomultipliers}

\author{S.~Korpar\address[Maribor]{Department of Chemistry and Chemical
    Engineering, University of Maribor, Maribor, Slovenia}\address[IJS]
  {Jo\v zef Stefan Institute, Ljubljana, Slovenia},
  I.~Adachi\address[KEK]{KEK, Tsukuba, Japan},
  H.~Chagani\addressmark[IJS]\thanks{Corresponding author, Email:
    hassan.chagani@ijs.si},
  R.~Dolenec\addressmark[IJS],
  K.~Hara\address[Nagoya]{Nagoya University, Nagoya, Japan},
  T.~Iijima\addressmark[Nagoya],
  P.~Kri\v zan\addressmark[IJS]\address[MPLJU]{Department of Mathematics and
    Physics, University of Ljubljana, Ljubljana, Slovenia},
  S.~Nishida\addressmark[KEK],
  R.~Pestotnik\addressmark[IJS]
  and
  A.~Stanovnik\addressmark[IJS]\address[EELJU]{Department of Electrical
    Engineering, University of Ljubljana, Ljubljana, Slovenia}}

\begin{document}

\begin{abstract}
A novel photon detector, the Silicon Photomultiplier (SiPM), has been tested in
proximity focusing Ring Imaging Cherenkov (RICH) counters that were exposed to
cosmic-ray particles in Ljubljana, and a 2~GeV electron beam at the KEK
research facility. This type of RICH detector is a candidate for the particle
identification detector upgrade of the BELLE detector at the KEK B-factory, for
which the use of SiPMs, microchannel plate photomultiplier tubes or hybrid
avalanche photodetectors, rather than traditional Photomultiplier Tubes (PMTs),
is essential due to the presence of high magnetic fields. In both experiments,
SiPMs are found to compare favourably with PMTs, with higher photon detection
rates per unit area. Through the use of hemispherical and truncated pyramid
light guides to concentrate photons onto the active surface area, the light
yield increases significantly. An estimate of the contribution to dark noise
from false coincidences between SiPMs in an array is also presented.
\end{abstract}

\maketitle

\section{INTRODUCTION}

Silicon Photomultipliers (SiPMs) are semiconductor photosensitive devices
consisting of an avalanche photodiode matrix on a common silicon substrate,
working in limited Geiger mode~\cite{Buzhan}. When compared with other position
sensitive detectors used in Ring Imaging Cherenkov (RICH) counters, they
possess the favourable property of insensitivity to high magnetic fields. They
operate at lower voltages than conventional photomultiplier tubes, have a high
peak photon detection efficiency that can be as high as 65\% at 400~nm, high
gain of~$10^{6}$ and good time response. Due to their small dimensions, they
allow compact, light and robust mechanical designs. These factors make them a
very promising candidate for a Cherenkov photon detector in a RICH counter.
However, due to the serious disadvantage of a very high dark rate
($\sim$~$10^{6}$~$\mbox{Hz/mm}^{2}$), they have not been considered until now
in such detectors where single photon detection is a major requirement.

One of the main goals of the present study is to verify the performance of
SiPMs as single photon detectors in a proximity focusing RICH
counter~\cite{Matsumoto04} proposed for the particle identification detector
upgrade of the BELLE detector at the KEK B-factory~\cite{Abe04}. An array of
silicon photomultipliers has been tested with Cherenkov photons from cosmic
muons. Different light guides have been machined and evaluated in an effort to
improve the efficiency of such a detector. Additionally, an $8\times 8$ array
of the new Surface Mounted Device (SMD) SiPMs coupled to individual light
guides has been exposed to a 2~GeV electron beam at the KEK research facility.
Finally, tests have been performed to estimate the contribution to dark noise
from false coincidence between SiPM channels.

\section{COSMIC-RAY TESTS}

The detection of Cherenkov light emitted by cosmic-ray particles traversing a
2.5~cm thick aerogel radiator of refractive index~1.045 is described in further
detail in~\cite{Korpar08}. The procedure and results are outlined briefly here.

An array of 12~Hamamatsu R5900-M16 Multianode Photomultiplier Tubes (MAPMTs)
and six Hamamatsu S10362-11-100U SiPMs of active surface area 1~$\mbox{mm}^{2}$
and pitch 100~$\mu\mbox{m}$~\cite{Hamamatsu} are mounted below the aerogel
radiator, and this entire set-up is contained within a light tight box. The
larger pixel size is preferred as the resultant increase in photon detection
efficiency outweighs the increased noise rate. The M16 MAPMTs have been used by
our group~\cite{Krizan97}, so their characteristics are well known and they
serve as a reference against which the parameters of the SiPMs are
investigated. An incident cosmic-ray particle passes through a scintillation
counter mounted above the box, providing a trigger signal. The path taken by
the particle is mapped by a series of three Multiwire Proportional Chambers
(MWPCs) located between the scintillator counter and the light tight box. The
arrival time and photon detector channel are recorded for each event.

A clear peak in the Cherenkov angle distribution of SiPM hits within a 3~ns
time window is observed. A factor of $5.4 \pm 0.2$ more photons are detected
per unit area by the SiPMs compared to the MAPMTs, in good agreement with the
expectation of~5.1 from the manufacturer~\cite{Hamamatsu}.

\section{IMPROVING THE PHOTON YIELD WITH LIGHT GUIDES}

The signal-to-noise ratio can be improved by increasing the number of hits per
single sensor. This can be achieved through the use of SiPMs with larger active
surface areas. However, this results in a marked increase in noise, and hence
diminishing returns.

\begin{figure}
  \begin{center}
    \includegraphics[width=7cm]{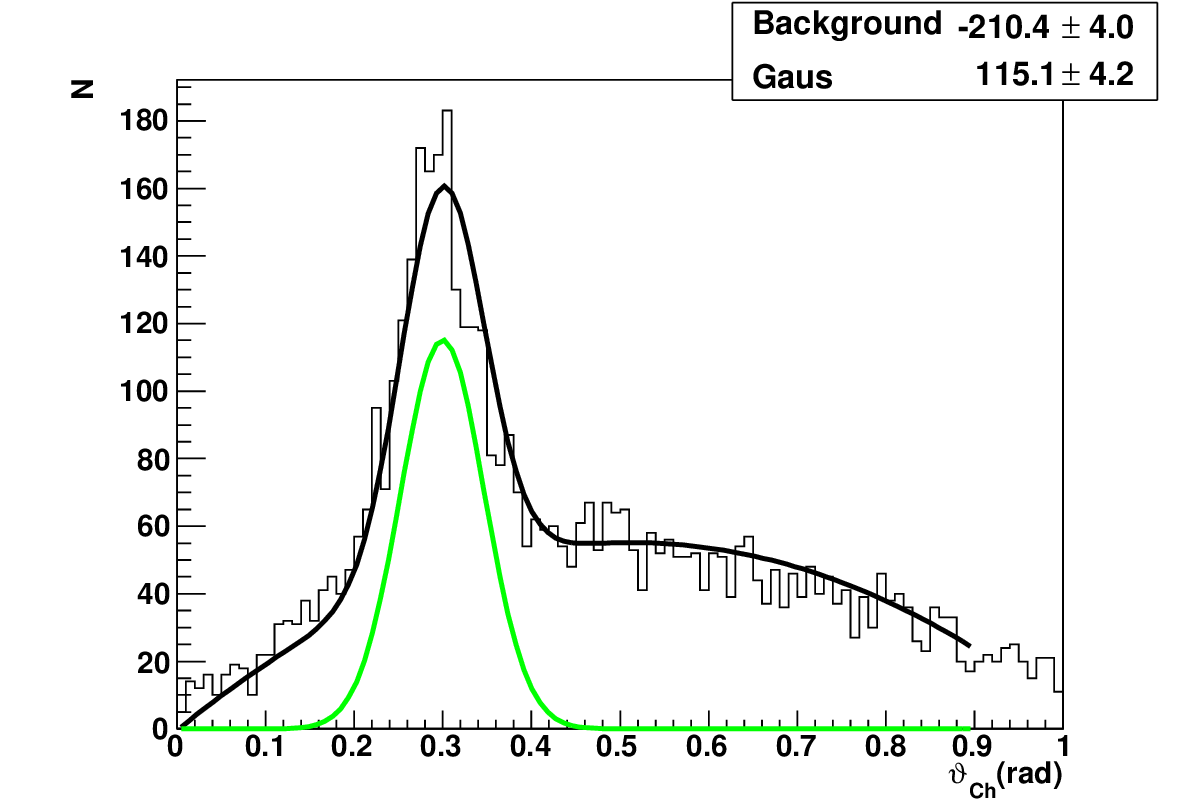}
    \vspace{-5mm}
    \caption{The distribution of SiPM hits from cosmic-ray tests, that are
      inside of the Cherenkov time window, as a function of the Cherenkov angle
      when hemispherical light guides are attached.}
    \label{cosmicresults}
  \end{center}
\end{figure}

Alternatively, the same result can be achieved by collecting light over a
larger area, and focusing it onto the smaller SiPM active surface with light
guides. The tops of blue light emitting diodes were attached to the six SiPMs
in the cosmic-ray set-up described above, and the procedure repeated, in an
attempt to ascertain the effects of hemispherical light guides. The resultant
Cherenkov angle distribution is shown in Fig.~\ref{cosmicresults}. A clear
improvement in the light yield by a factor of $3.6 \pm 0.2$ is witnessed, which
is in good agreement with the simulated value of~3.3. Further information on
the procedure and results is given in~\cite{Korpar08,Korpar08_2}.

Hemispherical light guides are suitable when the angle of incidence is limited
and the air gap between the SiPM active surface and the light collector is
large, as is the case in cosmic-ray set-up above. However, it is difficult to
machine a hemispherical light guide array for the 2~GeV electron beam tests at
KEK, and is unecessary if using a SiPM with a reduced epoxy protective layer,
such as the Hamamatsu S10362-11-100P Surface Mounted Device (SMD) with
1~$\mbox{mm}^{2}$ active surface area and 100~pixels~\cite{Hamamatsu}.
Therefore, a more suitable light collector, such as that in the shape of a
truncated pyramid, should be used when building a multi-channel module for this
purpose.

The geometry of the module, where each SiPM lies next to another, constrains
the maximum size of each light guide's entry surface to a square of side
2.54~mm, and the angle of inclination to $80^{\circ}$. By fixing these inherent
properties, ray-tracing simulations were performed where incoming photons were
distributed over a solid angle of $30^{\circ}$ to find an optimal light guide
length of 4~mm. An $8\times 8$ array of Hamamatsu S10362-11-100P SMD SiPMs with
a 0.3~mm protective layer were coupled to individual truncated pyramid light
guides that were machined from UV transparent perspex used in the HERA-B RICH
optical system~\cite{Broemmelsiek99}. Two-dimensional surface scans of the
module indicate an increase in light yield by a factor of~2.9 with the use of
these light guides.

\section{2~GeV ELECTRON BEAM TESTS}

\begin{figure}
  \begin{center}
    \includegraphics[width=5.5cm]{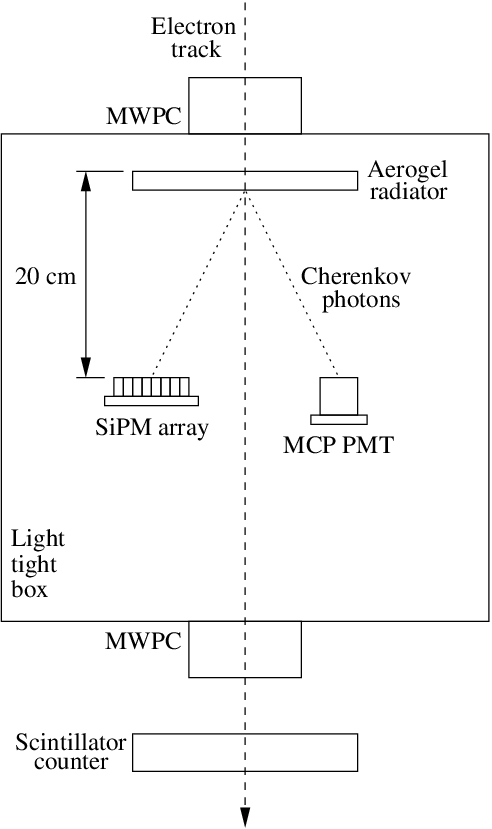}
    \vspace{-5mm}
    \caption{The experimental set-up for electron beam tests at KEK.}
    \label{beamtestschematic}
  \end{center}
\end{figure}

\begin{figure}
  \begin{center}
    \includegraphics[width=7cm]{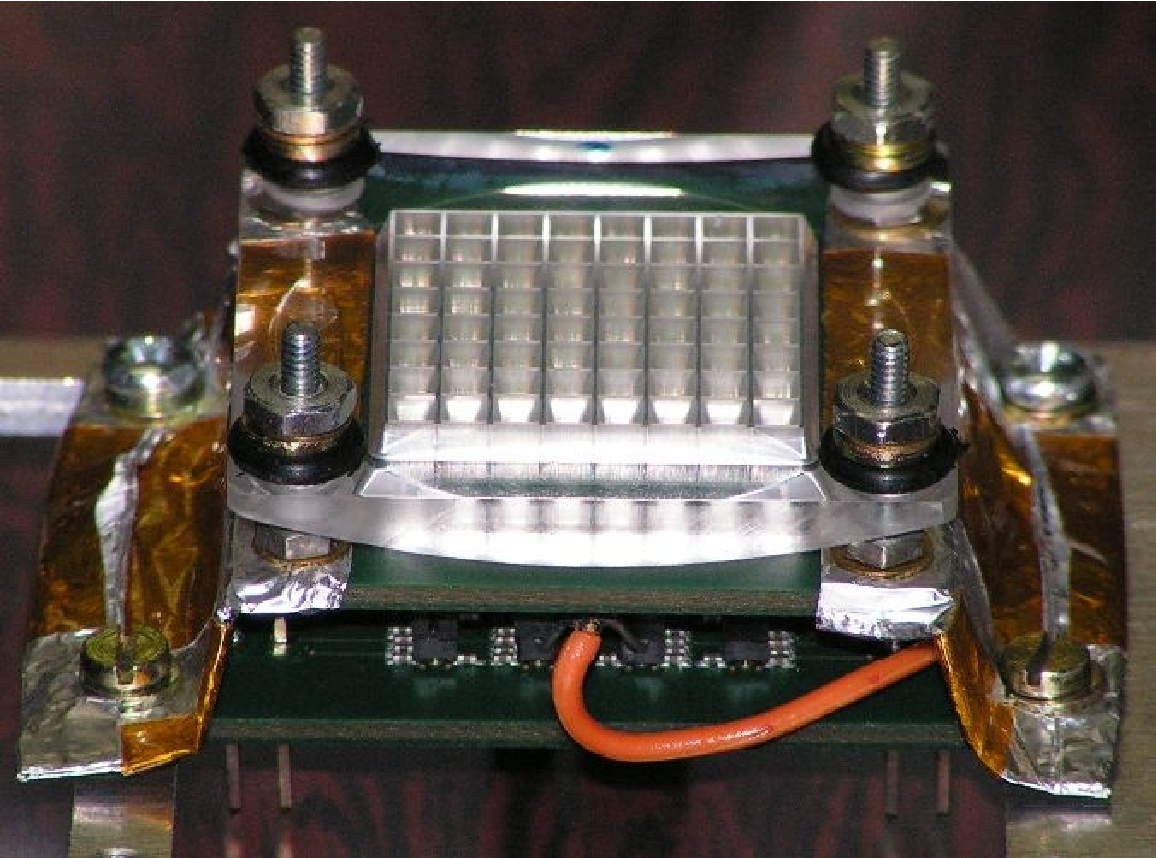}
    \vspace{-5mm}
    \caption{Truncated pyramid light guides attached to $8\times 8$ array of
      Hamamatsu S10362-11-100P SMD SiPMs used in beam tests at KEK.}
    \label{beamtestarray}
  \end{center}
\end{figure}

An outline of the experimental apparatus for the measurement of Cherenkov
photons is shown in Fig.~\ref{beamtestschematic}. Following the detection of an
incident electron by the plastic scintillator counter, its track coordinates
are obtained through delay line readout of the cathode plane signals from the
MWPCs located either side of a light tight box. The electron interacts with a
2~cm thick aerogel radiator of refractive index~1.045 contained within the box.
Cherenkov light is detected by the $8\times 8$ array of Hamamatsu
S10362-11-100P SMD SiPMs (Fig.~\ref{beamtestarray}) that lie 20~cm from the
radiator. Every $2\times 2$ block of SiPMs are added together to form a single
channel, resulting in a total of 16~SiPM channels. A Hamamatsu Microchannel
Plate Photomultiplier Tube (MCP-PMT), located the same distance away from the
aerogel radiator, acts as a reference for comparison with the parameters of the
SiPMs.

\begin{figure}
  \begin{center}
    $\begin{array}{c}
      \includegraphics[width=7cm]{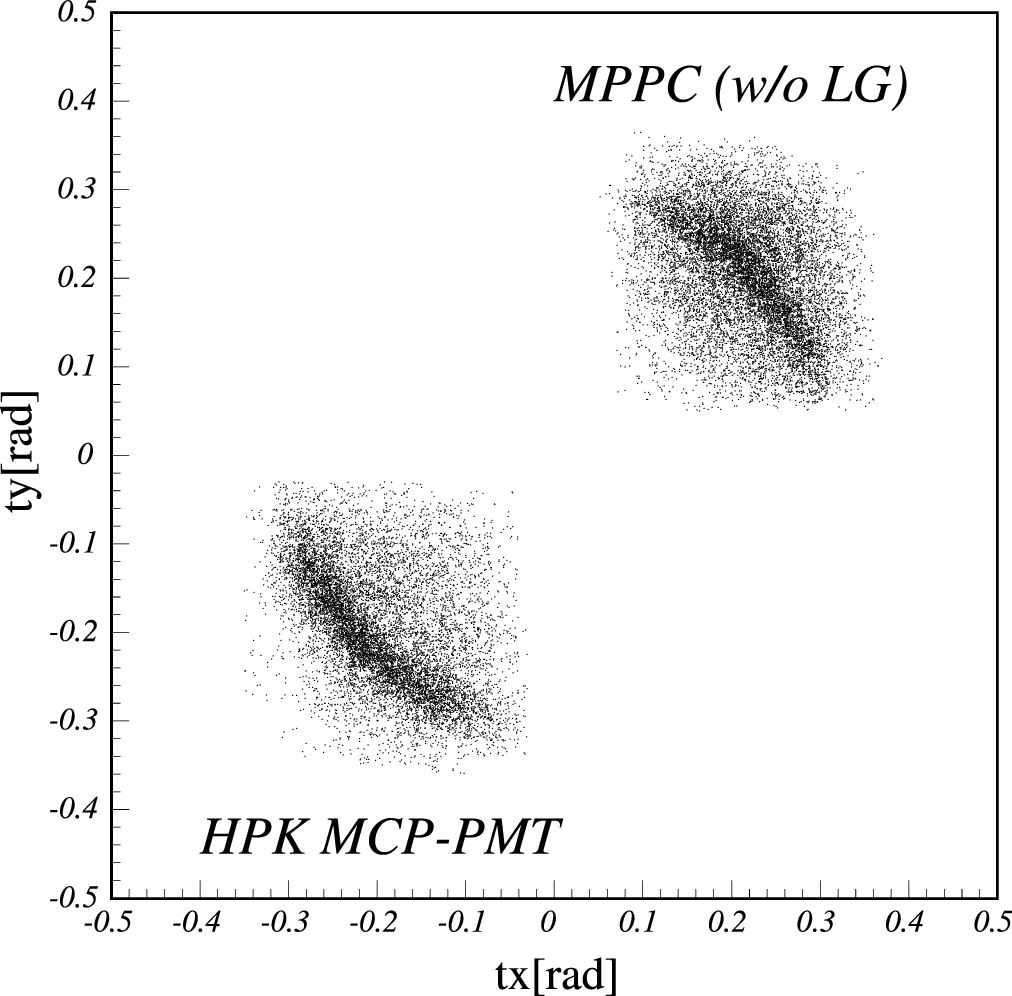} \\
      \includegraphics[width=7cm]{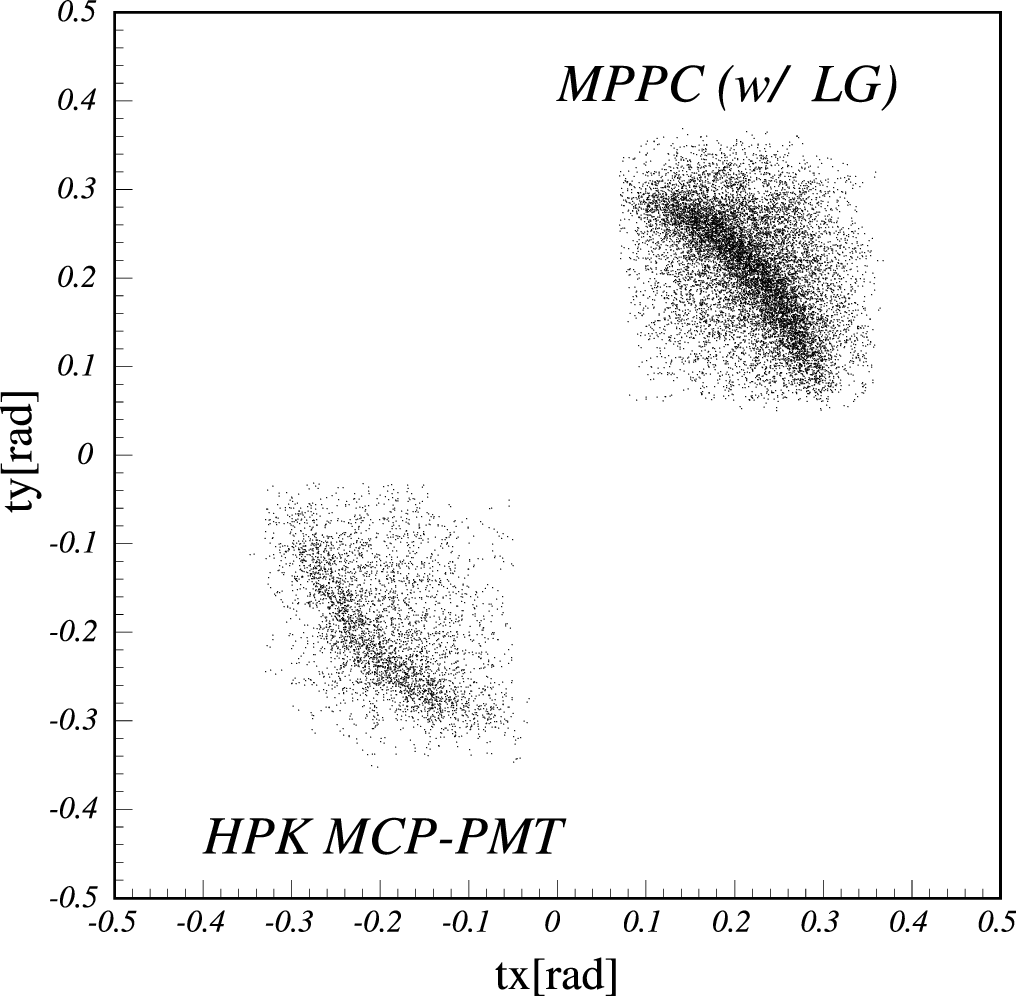}
    \end{array}$
    \vspace{-5mm}
    \caption{Hits in Cherenkov angle space without (top) and with (bottom)
      truncated pyramid light guides.}
    \label{cherenkovspace}
  \end{center}
\end{figure}

Tests have been performed without and with light guides attached to the SiPM
array. The signal-to-noise ratio improves by a factor of $\sim 2.7$ when light
concentrators are used. The distribution of hits in Cherenkov space is shown in
Fig.~\ref{cherenkovspace}. Due to the small size of the detector, the whole
Cherenkov ring is not shown. It is clear that there is a significant
improvement in the ratio of photons detected per unit area by the SiPMs to that
by the MCP-PMT through the use of light guides. Further analysis reveals that
this ratio is $\sim 1.1$ without light guides, and increases to $\sim 2.5$ when
the light guide array is attached.

\section{FALSE COINCIDENCE RATE}

In a proximity focusing RICH detector, the light emitted from Geiger discharge
in one SiPM, reflected off the radiator and detected by another is a source of
crosstalk between channels.

\begin{figure}
  \begin{center}
    \includegraphics[width=7cm]{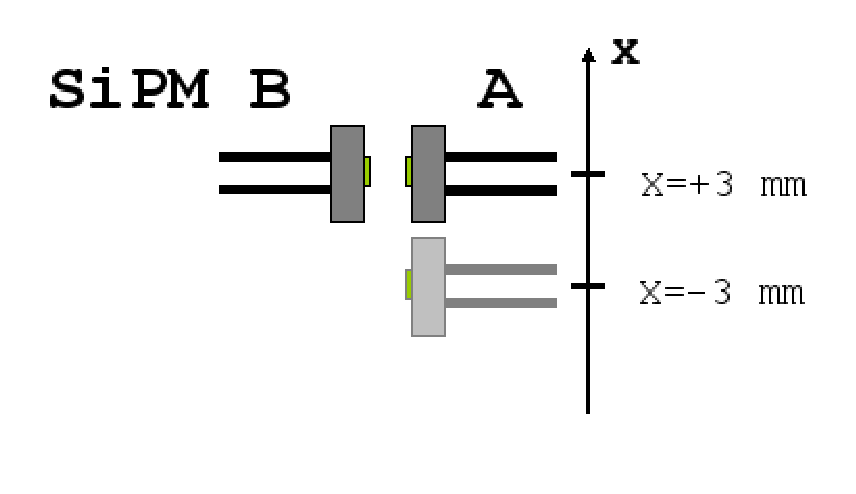}
    \vspace{-5mm}
    \caption{Experimental set-up to measure the optical crosstalk between a
      Hamamatsu S10362-11-100C SiPM (SiPM A) and a Hamamatsu S10362-11-050C
      SiPM (SiPM B).}
    \label{fcsetup}
  \end{center}
\end{figure}

The experimental set-up to measure the optical crosstalk, or false coincident
rate, between two SiPMs is shown in Fig.~\ref{fcsetup}. A Hamamatsu
S10362-11-100C SiPM (SiPM A), of 1~$\mbox{mm}^{2}$ active surface area,
100~pixels and 232~kHz dark rate at an operating voltage of 70~V, is placed on
a move-able stage. SiPM A faces a Hamamatsu S10362-11-050C SiPM (SiPM B), of
1~$\mbox{mm}^{2}$ active surface area, 400~pixels and 372~kHz dark rate at an
operating voltage of 71~V, a distance of 1~mm away. SiPM A moves perpendicular
to SiPM B, and they overlap when $x = 3$~mm.

\begin{figure}
  \begin{center}
    \includegraphics[width=7cm]{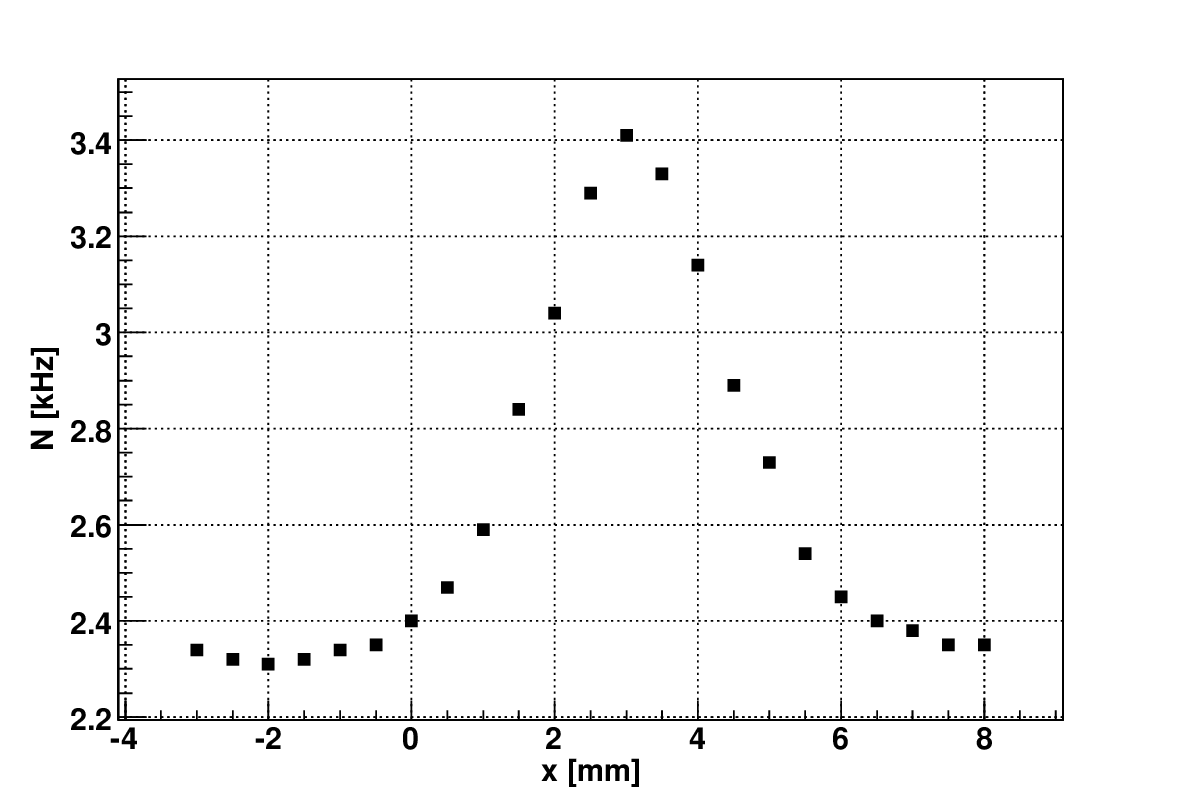}
    \vspace{-5mm}
    \caption{False coincidence rate as a function of distance $x$. The two
      SiPMs overlap when $x = 3$~mm.}
    \label{fcoincidence}
  \end{center}
\end{figure}

As shown in Fig.~\ref{fcoincidence}, the coincident dark noise rate at around
2.4~kHz, rises by approximately 1~kHz when the SiPMs overlap. For a planar
geometry, under conservative assumptions of reflectivity, this corresponds to
an increase of $\sim 0.1$\% in the dark count rate, which is negligible.

\section{CONCLUSIONS}

Single Cherenkov photons have been observed for the first time with SiPMs in a
RICH counter triggered by cosmic-rays. In an effort to improve the
signal-to-noise ratio, small light guides have been designed, manufactured and
tested. An $8\times 8$~SMD SiPM array coupled to individual light guides, which
were machined from UV transparent perspex used in HERA-B, has been tested in a
2~GeV electron beam at the KEK research facility. Tests of the false
coincidence rate between SiPMs indicates that this provides a negligible
contribution to the dark noise rate. Despite their relatively large dark noise,
SiPMs are promising Cherenkov photon detectors.

\section*{ACKNOWLEDGEMENTS}
This work was supported by the Slovenian Research Agency (ARRS), project
numbers J1-9340 and J1-9339, and a grant-in-aid for scientific research on
priority areas (``New Developments of Flavour Physics'', No. 18071003) from the
Ministry of Education, Culture, Sports, Science and Technology of Japan.

\end{document}